\documentclass{article}

\usepackage{arxiv}

\usepackage{cite}
\usepackage[utf8]{inputenc} 
\usepackage[T1]{fontenc}    
\usepackage{hyperref}       
\usepackage{url}            
\usepackage{booktabs}       
\usepackage{amsfonts}       
\usepackage{nicefrac}       
\usepackage{microtype}      
\usepackage{lipsum}
\usepackage{graphicx}
\usepackage{tabularx}
\usepackage{hyperref}
\usepackage{cleveref}
\usepackage{subcaption}

\graphicspath{ {./images/} }

\title{MemoryPods: Enhancing Asynchronous Communication in Extended Reality}

\author{
1\textsuperscript{st} Akos Nagy \\
\textit{Dept. of Networks \& Digital Media} \\
\textit{ECE, Kingston University}\\
Kingston upon Thames, UK \\
\texttt{A.Nagy@kingston.ac.uk}
\And
2\textsuperscript{nd} Yannis Spyridis \\
\textit{Dept. of Computer Science} \\
\textit{ECE, Kingston University}\\
Kingston upon Thames, UK \\
\texttt{Y.Spyridis@kingston.ac.uk}
\And
3\textsuperscript{rd} Gregory Mills \\
\textit{Dept. of Networks \& Digital Media} \\
\textit{ECE, Kingston University}\\
Kingston upon Thames, UK \\
\texttt{G.Mills@kingston.ac.uk}
\And
4\textsuperscript{th} Vasileios Argyriou \\
\textit{Dept. of Networks \& Digital Media} \\
\textit{ECE, Kingston University}\\
Kingston upon Thames, UK \\
\texttt{Vasileios.Argyriou@kingston.ac.uk}
}

\begin{document}
\maketitle

\begin{abstract}
    Asynchronous communication has become increasingly essential in the context of extended reality (XR), enabling users to interact and share information immersively without the constraints of simultaneous engagement. However, current XR systems often struggle to support effective asynchronous interactions, mainly due to limitations in contextual replay and navigation. This paper aims to address these limitations by introducing a novel system that enhances asynchronous communication in XR through the concept of MemoryPods, which allow users to record, annotate, and replay interactions with spatial and temporal accuracy. MemoryPods also feature AI-driven summarisation to ease cognitive load. A user evaluation conducted in a remote maintenance scenario demonstrated significant improvements in comprehension, highlighting the system’s potential to transform collaboration in XR. The findings suggest broad applicability of the proposed system across various domains, including direct messaging, healthcare, education, remote collaboration, and training, offering a promising solution to the complexities of asynchronous communication in immersive environments.
\end{abstract}

\section{Introduction}

Asynchronous communication has recently emerged as a pivotal mechanism in the realm of extended reality (XR), allowing users to interact and exchange information without the need for simultaneous engagement. This mode of communication is crucial in XR environments, where users frequently participate in immersive experiences across various time zones or working hours, thus transcending traditional temporal and spatial boundaries. In these contexts, effective collaboration relies on support for asynchrony, which offers distinct benefits compared to synchronous communication, including parallel work capabilities, flexible time coordination, promotion of engagement \cite{liaqat2024promoting}, and opportunities for reflection \cite{hollan1992beyond,olson2000distance}. The growing relevance of XR in domains such as education \cite{pregowska2024will,bruuvza2021vrdeo}, training \cite{buttner2020augmented, palmas2020defining}, remote collaboration \cite{burova2022asynchronous,pereira2019extended}, and healthcare \cite{shaikh2022data} highlights the need for effective asynchronous communication systems in this context. In such environments, users benefit from being able to review and contribute to interactions at their convenience, which can enhance communication effectiveness, facilitate more flexible collaboration, and optimise the use of time and resources. However, the immersive and dynamic nature of XR introduces unique challenges that make it difficult to implement effective asynchronous communication solutions \cite{chow2019challenges}.

Current XR systems often encounter challenges in supporting the required functionalities for effective asynchronous capabilities. A primary limitation is the limited development of contextual replay options, which makes it difficult for users to navigate and comprehend past events within the XR environment efficiently. Conventional replay mechanisms are often linear, restrictive, and do not fully capture the interactive and spatial dynamics of XR interactions. Furthermore, the tools available for annotating events are often underdeveloped, making it challenging for users to pinpoint critical moments and derive actionable insights. These challenges limit the potential of XR applications, particularly in situations where understanding the details of previous interactions is crucial.

In addition, navigating through extensive, unindexed recordings leads to wasting time and increasing cognitive load for the user. This inefficiency hinders the ability to extract meaningful insights and make informed decisions. Additionally, the absence of automated summarisation features can lead to missed opportunities or misunderstandings, further limiting the value of asynchronous communication. To address these challenges, there is a clear need for systems that enable users to quickly locate, review, and understand critical moments within asynchronous XR recordings, thereby enhancing their understanding and decision-making capabilities. Such systems can significantly enhance the user experience by providing precise, context-rich, and user-friendly tools for reviewing and understanding sequential events \cite{lilija2020put}.

In light of the above, the primary focus of this paper is to introduce a novel system which revolves around the concept of \textit{MemoryPods}, a feature specifically designed to enhance asynchronous communication within immersive environments. MemoryPods are interactive digital containers that store comprehensive records of past events, including spatial, visual, and audio data. They allow users to revisit and analyse these moments from multiple perspectives, offering an immersive replay experience. MemoryPods also include intuitive annotation features, enabling users to mark and easily navigate key events, ensuring that important interactions can be reviewed efficiently. This is particularly useful for collaborative remote maintenance, where team members often need to review and coordinate tasks asynchronously, allowing for effective communication and analysis without the need for real-time presence. Central to the proposed system are the following key contributions:

\begin{itemize}
    \item MemoryPod: A versatile mechanism enabling users to record and revisit past events with adaptable, multi-perspective options. This system allows for the replay of scenarios either in the original real-world environment where the event occurred, given that the user is physically present, or within their own  space, using miniature scale options. This flexibility facilitates a comprehensive analysis of recorded interactions, accommodating different spatial contexts and scales.
    
    \item Spatio-temporal Annotations: Strategic markers indicating critical moments in the recorded events, facilitating efficient navigation and retrieval of pivotal information.
    
    \item AI-Driven Summarisation: Large language model (LLM) incorporation to provide automatic concise summaries of event transcripts in real-time, highlighting key details and allowing users to quickly grasp the essential aspects of complex scenarios.

    \item Modular AI Integration: A system architecture designed to seamlessly integrate various AI tools, thereby enhancing the analytical capabilities and user experience in reviewing and interpreting recorded XR interactions.
\end{itemize}

To assess the effectiveness of the proposed system, a user evaluation was conducted in a collaborative remote maintenance process. Participants were tasked with reviewing recorded interactions and recounting key aspects of the procedure, such as timing, duration, and spatial context. Results demonstrated a significant improvement in user comprehension when using the proposed system, suggesting the effectiveness of the introduced contributions in enhancing asynchronous communication in XR.

By addressing some of the complexities of communication in XR environments, the proposed system offers a promising solution for improving collaboration, learning, and decision-making within different XR contexts. The system’s potential applications extend across various domains, including education, remote collaboration, and virtual training. For instance, in education, it can reform how instructors and students engage with content, offering a more interactive and insightful learning experience. In remote collaboration, it can improve team dynamics by providing precise tools for analysing past meetings. Additionally, in virtual training it can transform how trainees review and learn from scenarios, leading to more effective outcomes.

The rest of the paper is structured as follows: Section 2 reviews the existing literature on asynchronous communications in extended reality and associated challenges. Section 3 describes the key components of the proposed system, and outlines the operational framework, detailing the functionality of MemoryPods. Section 4 explains the procedure, measures, and results of the conducted empirical study, followed by a discussion of the findings. Lastly, Section 5 concludes the paper by summarising the contributions and potential directions for future research.

\section{Related Work}

\subsection{Extended reality for remote collaboration}

Recent advancements in extended reality have demonstrated significant potential for enhancing remote collaboration, particularly in asynchronous settings. Several key systems have explored how immersive technologies can improve knowledge transfer and interaction between users working in different locations and times. VRdeo \cite{bruuvza2021vrdeo} is an educational tool that allows teachers and students to interact asynchronously through recorded virtual reality (VR) sessions, which can be exported as videos. Creators can teleport around the scene, manipulate 3D objects, and provide voice instructions during recording. A user study comparing VR and 2D recordings found that participants preferred VR for its interactivity and engagement, resulting in improved learning outcomes.

Providing a structured way to access and learn from previous activities can significantly enhance collaboration in XR settings. The Tesseract system \cite{mahadevan2023tesseract} improves remote collaboration in spatial design processes, by enabling users to capture and review design recordings, and allowing teams to reflect on workflows and design decisions asynchronously. Its Worlds-in-Miniature-based Search Cube interface lets users stage and query parts of a design, making it easier to retrieve key moments from past sessions. This feature aids knowledge transfer, as junior designers can replay senior designers' work to understand design rationale. 

\subsection{Enhancing user experience in asynchronous extended reality}

Several works have explored innovative methods for enhancing user interaction and understanding in asynchronous XR environments. RealityReplay \cite{cho2023realityreplay} introduces an end-to-end pipeline that detects and visualises important changes in a user's environment through user-centric sensing. The system combines semantic segmentation and saliency prediction to track significant changes in the environment, filtering out static or irrelevant objects. Users can select an area of interest, and the system generates a summary visualisation, offering options from simple motion lines to more detailed depictions of texture and position. Who Put That There? \cite{lilija2020put} is a VR system designed to track spatio-temporal changes in objects, allowing users to query a timeline of changes such as location, size, and appearance. It introduces interaction primitives, that include a "trajectory sphere" for exploring object interactions over time and a "preview sphere" that isolates specific events without altering the surrounding virtual environment. An evaluation study showed that users preferred the enhanced navigation tools over conventional timelines, noting increased engagement and a stronger sense of presence.

Spatio-temporal tools can significantly enhance asynchronous learning by helping to lower cognitive load. The XR-LIVE system \cite{thanyadit2022xr} facilitates this through an assistive toolkit that includes features like a checklist for organising lab tasks and an auto-pause function that halts demonstrations at key moments. Temporal cues guide learners through each step, while spatial features allow students to accurately replicate instructor actions within a virtual environment, facilitating a deeper understanding of complex tasks. Together, these elements support different learning styles, boosting engagement and making asynchronous learning more effective.

\subsection{Design challenges in asynchronous collaboration}

While there is a relative scarcity of research for asynchronous collaboration in XR, the current literature presents distinct design challenges, indicating the field is still developing. A key challenge lies in the lack of established methods for representing changes made in XR environments to users who are not present simultaneously, which is crucial for maintaining continuity in tasks \cite{pidel2020collaboration}. Additionally, asynchronous tools can contribute to information overload, especially when users navigate multiple communication channels. 

Asynchronous collaboration involves more than simply producing and consuming information; it requires precise authoring of spatial annotations that are contextually and temporally accurate \cite{irlitti2016challenges}. Key challenges include creating annotations that reflect both location and time, allowing users to understand the sequence and significance of events. In particular, in remote maintenance, asynchronous collaboration should enable experts to guide on-site technicians through the placement of spatially aligned annotations, that provide clear instructions for equipment repair and servicing \cite{marques2021remote}. In addition, asynchronous AR systems should support user-driven behaviours, such as gaze tracking, and facilitate collaboration across different devices and environments, enhancing the depth and efficiency of communication.

Building on these challenges, the integration of privacy, security, and user data protection in XR collaboration further complicates the design of asynchronous systems. Effective access control is crucial in XR environments to manage permissions and prevent unauthorised access to shared virtual content and physical spaces \cite{rajaram2023eliciting}. Privacy concerns arise from environmental sensing, where unintended capture of users' surroundings can occur without consent. In this context, balancing security with usability is essential, as complex interaction techniques may enhance security but complicate the user experience. 

\section{System Design}

\begin{figure}[b!]
  \centering  
  \includegraphics[width=0.7\linewidth]{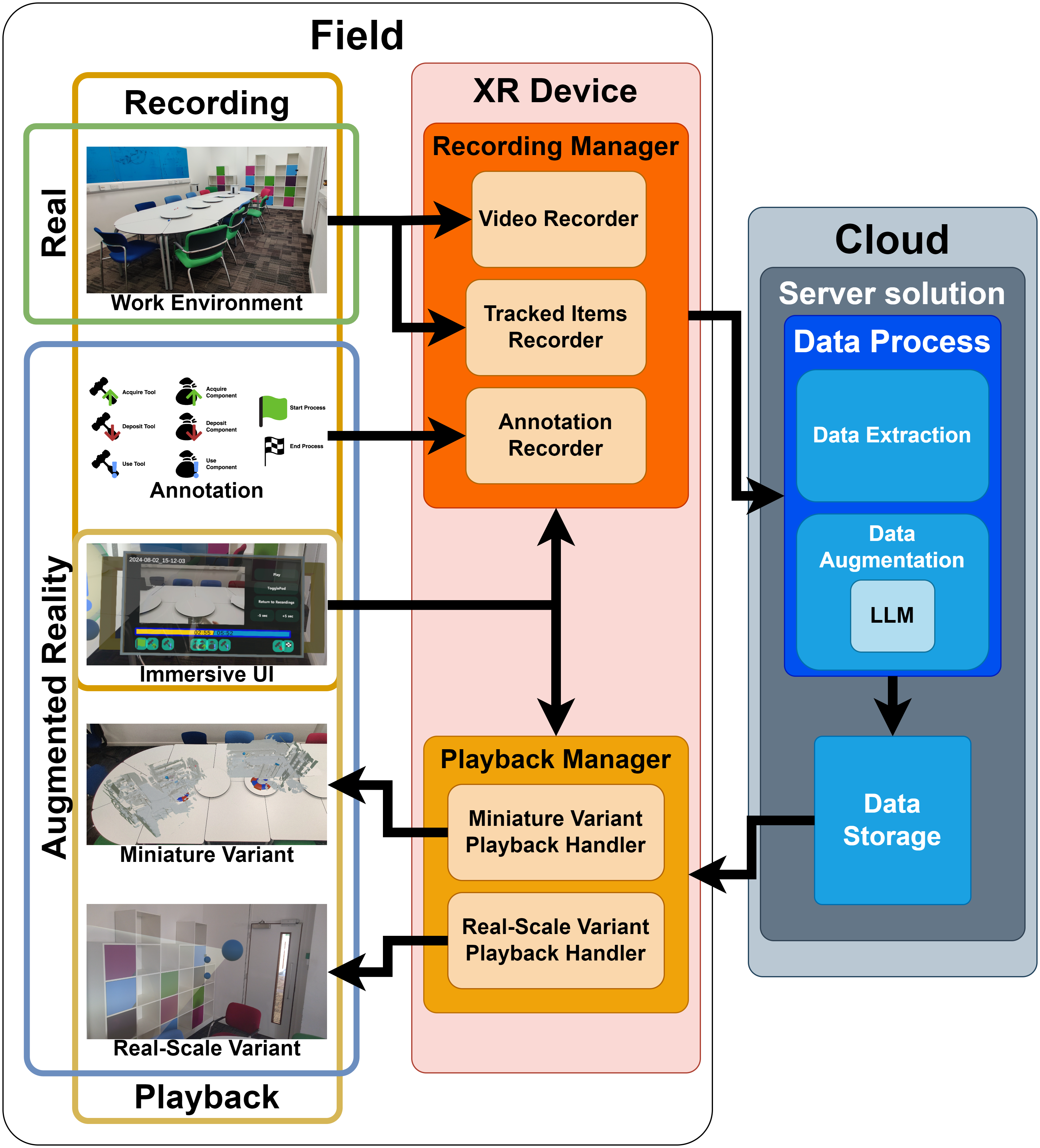}
  \caption{Workflow of the proposed system for enhancing asynchronous communication in XR environments. The process begins with real-world data recording through an XR device, capturing video, tracked items, and annotations. The data is sent to a cloud server for processing, where LLMs aid in data extraction and augmentation. Playback is offered in two forms: a miniature variant or a real-scale variant, both accessible in augmented reality.}
  \label{fig:workflow}
\end{figure}

The proposed system, is designed to enhance asynchronous communication and collaboration in XR, with particular emphasis on remote maintenance scenarios. By utilising the MemoryPods - recording units that capture and store detailed interactions and events - the system allows users to record, annotate, and replay complex real-world events within XR environments. The system architecture integrates an XR device for capturing video, spatial data, and user-defined annotations during live interactions. This data is processed on a cloud server, where LLMs augment the captured information by extracting key insights and generating summaries.

Two playback variants are offered for reviewing MemoryPods: a miniature-scale mode that allows users to review events in a compact, spatially optimised format, and a real-scale mode for more immersive, contextually accurate analysis. Both options support dynamic replays, enabling users to revisit critical moments of past interactions. The system’s flexible, modular design allows seamless integration into various remote collaboration and maintenance workflows, providing users with intuitive tools to efficiently review and interact with MemoryPods. The workflow of the system is illustrated in Figure \ref{fig:workflow}.

\subsection{System components}

\subsubsection{Contextual annotations}
\begin{figure}[t!]
\begin{subfigure}{.3\textwidth}
  \centering
  \includegraphics[width=.95\linewidth]{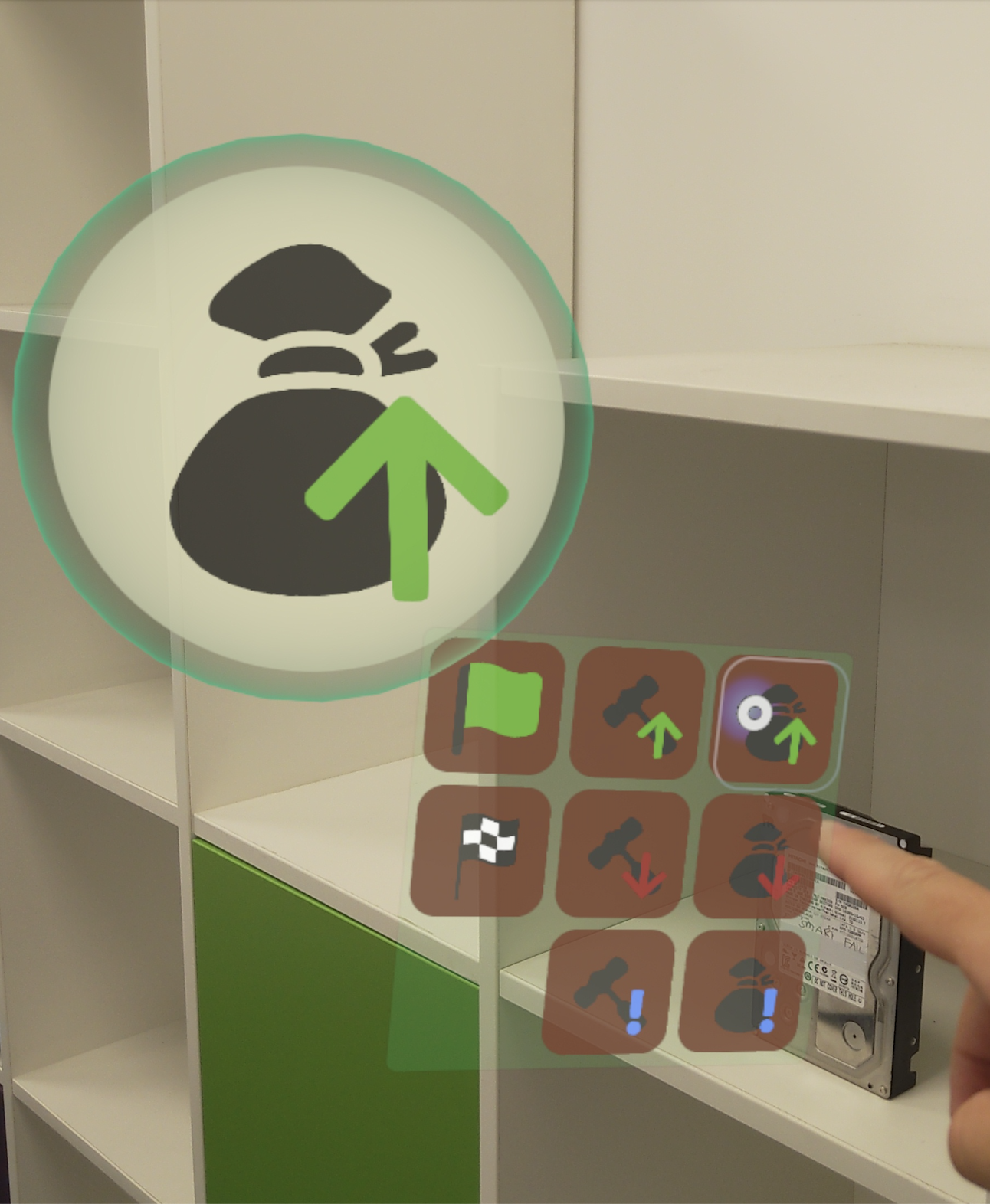}
  \caption{Acquiring annotation.}
  \label{fig:sfig1}
\end{subfigure}%
\begin{subfigure}{.6\textwidth}
  \centering
  \includegraphics[width=.95\linewidth]{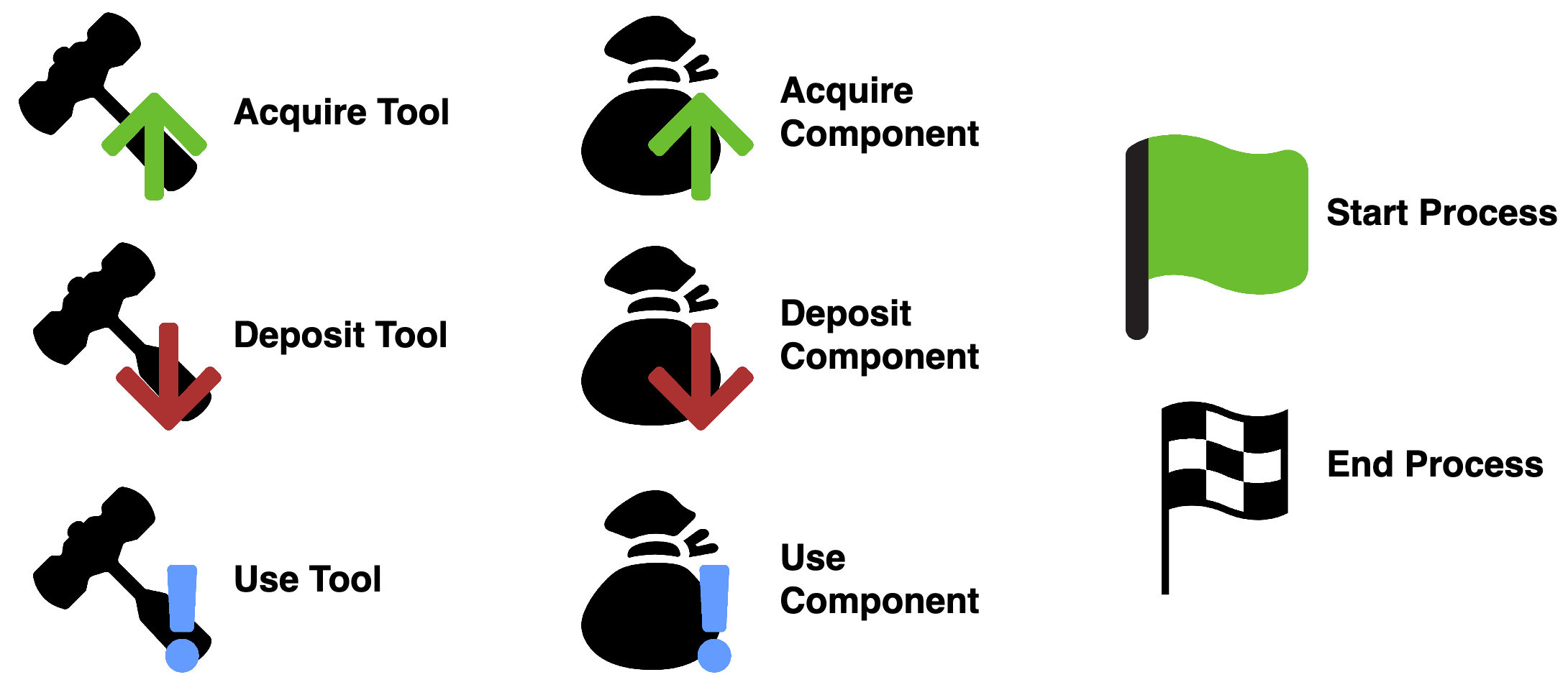}
  \vspace{2em}
  \caption{List of the available annotations and their description.}
  \label{fig:sfig2}
\end{subfigure}
\caption{Annotation procedure. The user places a spatio-temporal annotation to indicate when and where this component should be acquired.}
\label{fig:annotations}
\end{figure}

Contextual annotations constitute a vital component of the system by providing visual markers that highlight important locations or actions within the XR environment. These annotations deliver essential contextual information directly into the augmented space. They act as visual cues to enhance user awareness and guide interactions, marking critical areas or events that are of particular interest. This feature is especially beneficial in complex scenarios where spatial context and timing are important for a thorough understanding. These annotations ensure that critical information is easily indicated during the initial experience and equally accessible in subsequent replays, enhancing the efficiency of the communication by offering visual references to pivotal moments or locations. The system supports annotating actions for acquiring, using, and depositing tools or components related to the maintenance procedure, as well as indications for starting and ending the process, as demonstrated in Figure \ref{fig:annotations}.

\subsubsection{Spatial Anchor Point}
An important aspect of the proposed system is the spatial anchor point, which serves as the foundational framework for recording, storing, and replaying events with precise spatial fidelity. This feature utilises a calibration marker (e.g., a QR code) strategically placed within the real-world environment, to act as a reference point. The spatial anchor is calibrated with the XR headset to ensure accurate positional and orientational data. Through the integration of the anchor point, the system captures and maintains the exact spatial relationship between the real-world environment and the virtual elements. This ensures that when events are replayed, they retain their original spatial context, whether the user is physically present in the initial environment or in a different location, thus providing a robust mechanism for spatial coherence. 

\subsubsection{Body Movement Tracking}
The system has the ability to track body movement accurately, through embedded capabilities within modern XR headsets, which come equipped with integrated sensors and cameras for monitoring the trajectories and movements of various body segments. Such headsets continuously capture dynamic positional data of critical body parts, such as hands, head, and limbs, and feed this information into the system's spatial reference framework. By synchronising this movement data with the spatial anchor point, the system ensures that virtual elements are precisely aligned with user actions. This seamless integration enhances the interactive experience, providing a high level of spatial coherence and responsiveness within the augmented environment. 

This feature also aims to improve context awareness during the replay of XR recordings. By capturing and replaying movements with high fidelity, the system allows users to revisit past interactions in a manner that closely mirrors the original experience. This is particularly beneficial in scenarios where understanding the spatial and bodily context is crucial, such as in those related to remote maintenance environments. The precise synchronisation of movement data with the recorded environment ensures that users can effectively analyse and interpret past actions, leading to a deeper understanding of the events and interactions, and therefore a clear and immersive perspective of the recorded scenario.

\subsubsection{Environment recording}
\begin{figure}[t!]
\begin{subfigure}{.5\textwidth}
  \centering
  \includegraphics[width=.95\linewidth]{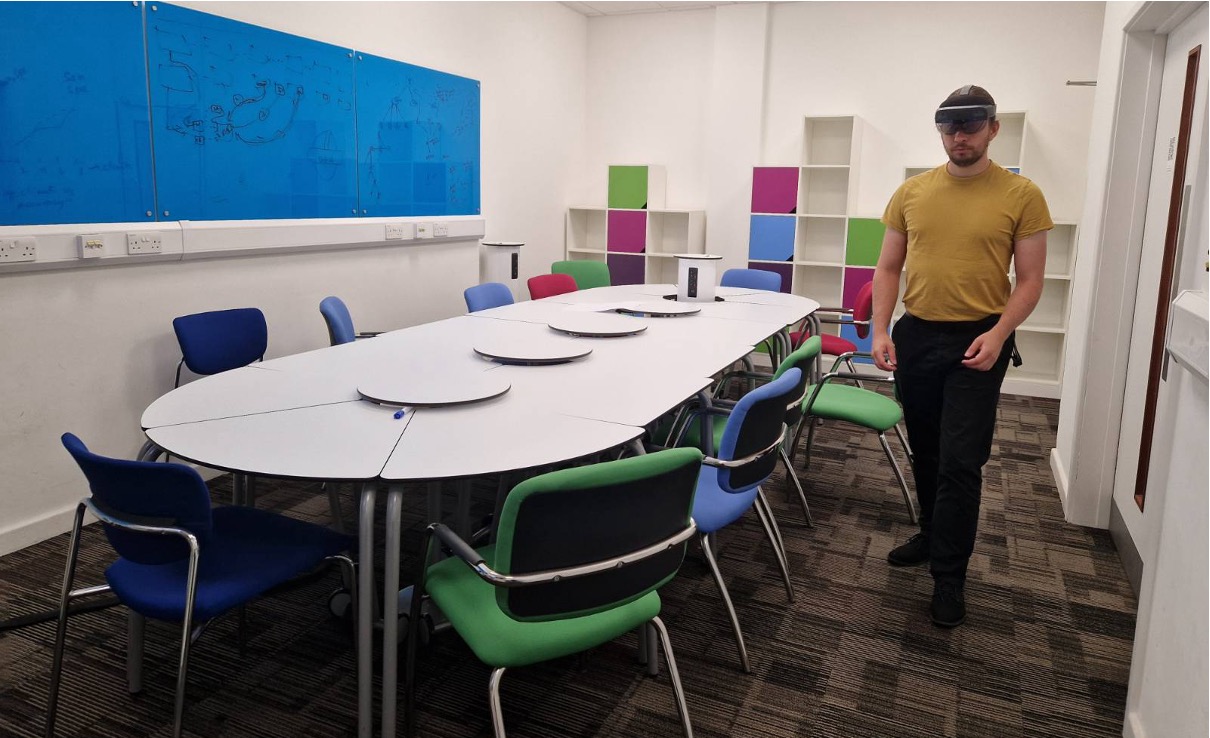}
  \caption{Original Environment.}
  \label{fig:reconstruction_original}

\end{subfigure}%
\begin{subfigure}{.5\textwidth}
  \centering
  \includegraphics[width=.95\linewidth]{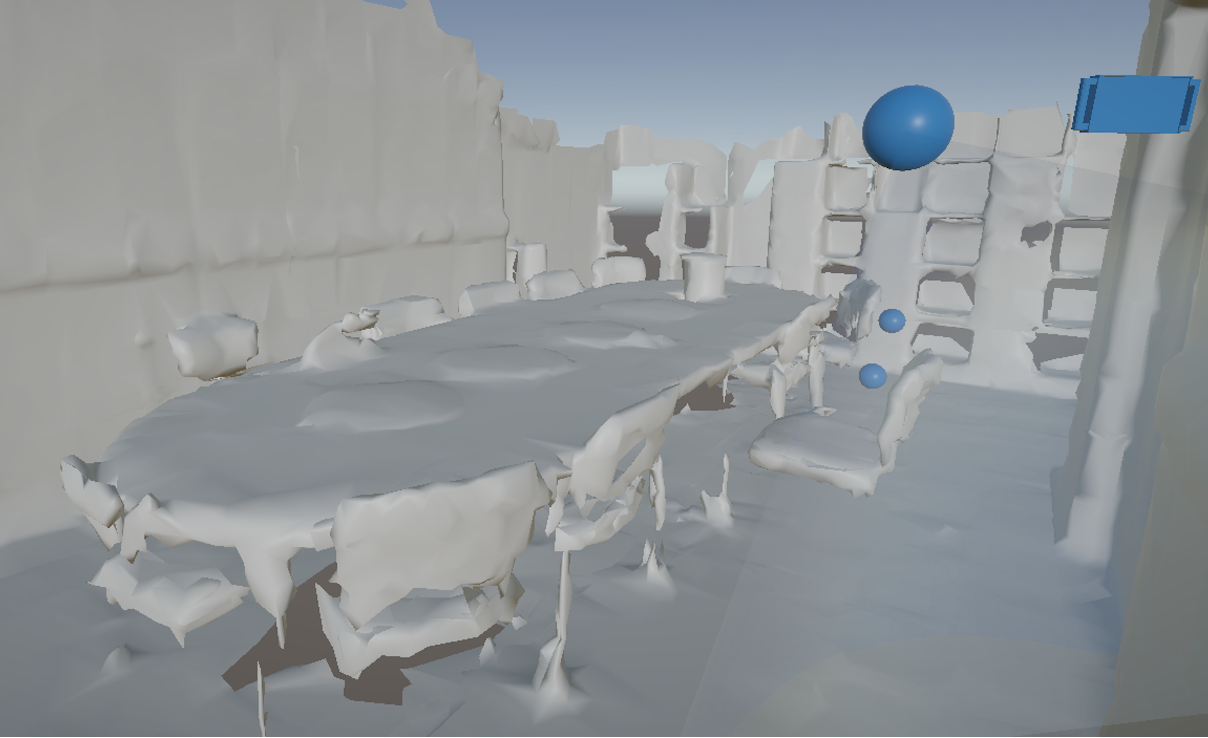}
  \caption{3D Reconstruction}
  \label{fig:reconstruction_virtual}
\end{subfigure}
\caption{Original environment and the 3D Reconstruction of its recording. The large blue sphere represents the head of the user, the smaller spheres indicate their hands, while the faded triangle showcases their field of view.}
\label{fig:reconstruction}
\end{figure}

The system is capable of capturing details of the physical surroundings, which is crucial for generating comprehensive three-dimensional (3D) mesh data. The process involves capturing the spatial characteristics of the environment, including various surfaces, structures, and objects, and translating this information into a high-fidelity 3D mesh representation, as demonstrated in Figure \ref{fig:reconstruction}. Such detailed environmental modelling ensures that virtual objects can be integrated with high spatial accuracy, thus maintaining the coherence and realism of the augmented experience. This precise environmental capture allows users to interact with virtual elements that seamlessly blend into the physical space, enhancing the overall immersion and usability of the system in a practical, real-world context. 

The detailed environmental capture is directly utilised to create the miniature scale variant of the MemoryPod. By accurately representing the environment in a 3D mesh, the system can scale down the spatial data while preserving the relative positions and interactions of virtual objects. This allows for the generation of a miniature version of the scene that maintains the spatial integrity of the original setup.

\subsubsection{Narrative Abstraction}
\begin{figure}[t!]
  \centering  
  \includegraphics[width=0.8\linewidth]{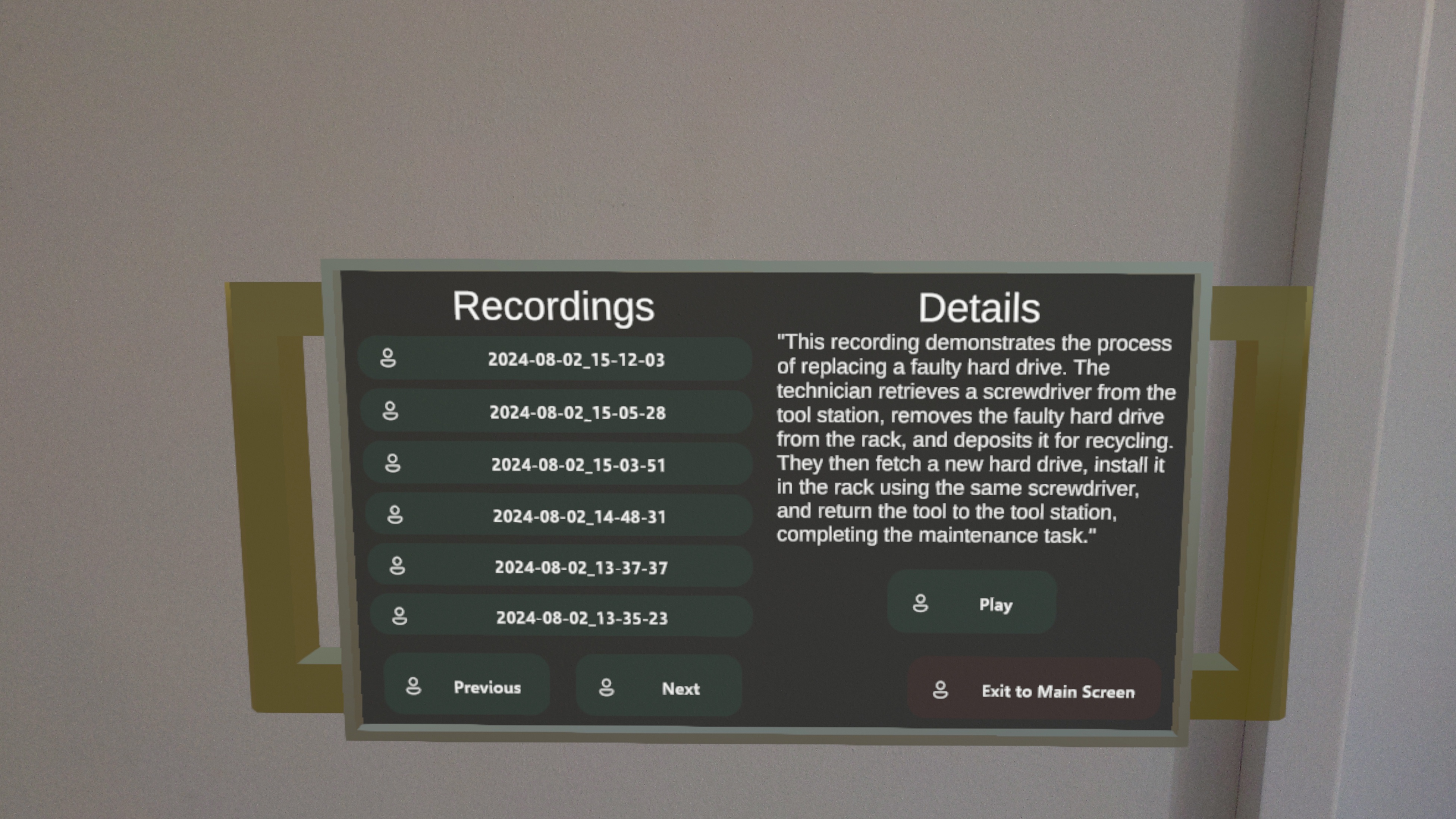}
  \caption{Narrative abstraction of the selected recording using LLMs.}
  \label{fig:summarisation}
\end{figure}

This system leverages LLMs to generate concise, contextually rich summaries of recorded events. By extracting and distilling key information from extensive event logs, the summarisation component provides users with a quick overview of critical moments, significantly reducing the cognitive load and time required for thorough event review. The AI-generated summaries are then presented in a user-friendly format, highlighting essential details such as key decisions, actions, and environmental changes, which allows users to rapidly grasp the core elements of more complex scenarios. This feature greatly enhances the efficiency of event review and also supports a deeper understanding of the context and flow of events. Users can rely on these summaries to quickly orient themselves within the broader narrative of the recorded interaction, making it easier to revisit specific moments or gain insights into the overall dynamics of the event. An example of narrative abstraction is depicted in Figure \ref{fig:summarisation}.

\subsection{Operational Framework}

Aligned with the Industry 5.0 human-centric approach, the system is designed with an intuitive, user-friendly interface centred around a virtual device, illustrated in Figure \ref{fig:summarisation}, that enables all the system's functionalities. This interface is seamlessly integrated into the XR scene and is modelled to visually resemble a real-world device, a design aiming to minimise visual obstruction while providing users with familiar visual cues, thereby enhancing their interaction within the XR environment. For example, the device features prominently designed handles that intuitively indicate contact points, allowing users to easily manipulate it and interact with the scene. This approach facilitates natural and efficient user interactions, allowing straightforward event recording and replay processes.

\subsubsection{Event Recording}
Upon initialisation of the session, the system identifies and records the position of the anchor point in the physical environment, which acts as the reference for this session, thus allowing the system to determine the precise coordinates in the physical space. Once the anchor point is established, users can start introducing event indicators into the XR scene, through engagement with the virtual device. For example, in a collaborative remote maintenance scenario, users might add annotations to machinery components, or tools within the physical space, and indicate them as key objects. The system computes the spatial coordinates of each object with respect to this origin, enabling the calculation of relative distances, angles, and orientations. Additionally, the system records audio and visual data throughout the session, capturing both environmental sounds and visual interactions. This capability allows users to collaboratively identify and address maintenance issues by visually marking problem areas and providing detailed instructions directly within the virtual scene.

During the event recording, the system logs spatial parameters alongside the temporal sequence of interactions, creating a comprehensive record of the event inside the MemoryPod, including the contextual annotations placed to the objects or the scene.  These representations capture both the spatial coordinates and the precise timing within the recording timeframe, and are stored in a structured format that enables efficient retrieval and analysis.

\subsubsection{Event Replay}

For event replay, the system first re-establishes the reference point by detecting the anchor point in the physical environment. Once the anchor point is confirmed, the application retrieves the recorded data and reconstructs the scene, ensuring that all annotations are accurately placed according to their original spatial and temporal context. This process guarantees that the spatial layout and interactions are preserved, providing the user with an accurate and authentic reproduction of the originally recorded event.

\begin{figure}[t!]
  \centering  
  \includegraphics[width=0.8\linewidth]{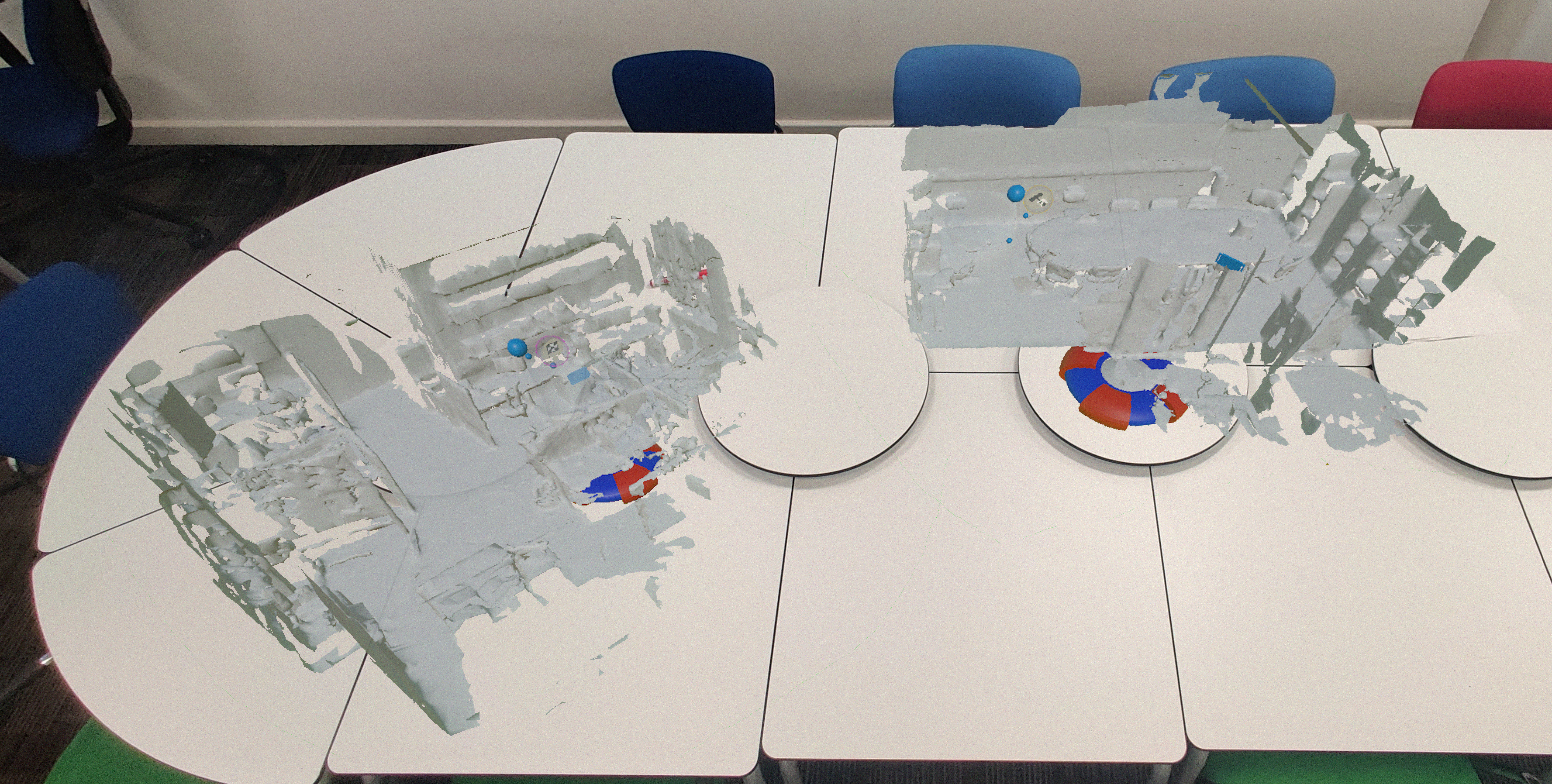}
  \caption{Illustration of two XR recordings in their miniature scale version, placed simultaneously within the environment. This feature enables users to review and manage multiple recorded processes concurrently, facilitating multitasking within the immersive space.}
  \label{fig:miniature_variants}
\end{figure}

In case the user is not present at the original physical environment, the system supports the miniature scale variant. In this mode, the system scales down the spatial data and reconstructs the scene in a compact, miniature format. Users can place this scaled-down version on any flat surface, such as a table, enabling an overview of the entire event. Despite the reduction in scale, the system maintains the integrity of the spatial and temporal relationships captured during the original session, ensuring that users can still conduct detailed analysis of the recorded interactions. In addition, multiple miniature variants can be placed and viewed simultaneously, as depicted in Figure \ref{fig:miniature_variants}.

\begin{figure}[t!]
  \centering  
  \includegraphics[width=0.8\linewidth]{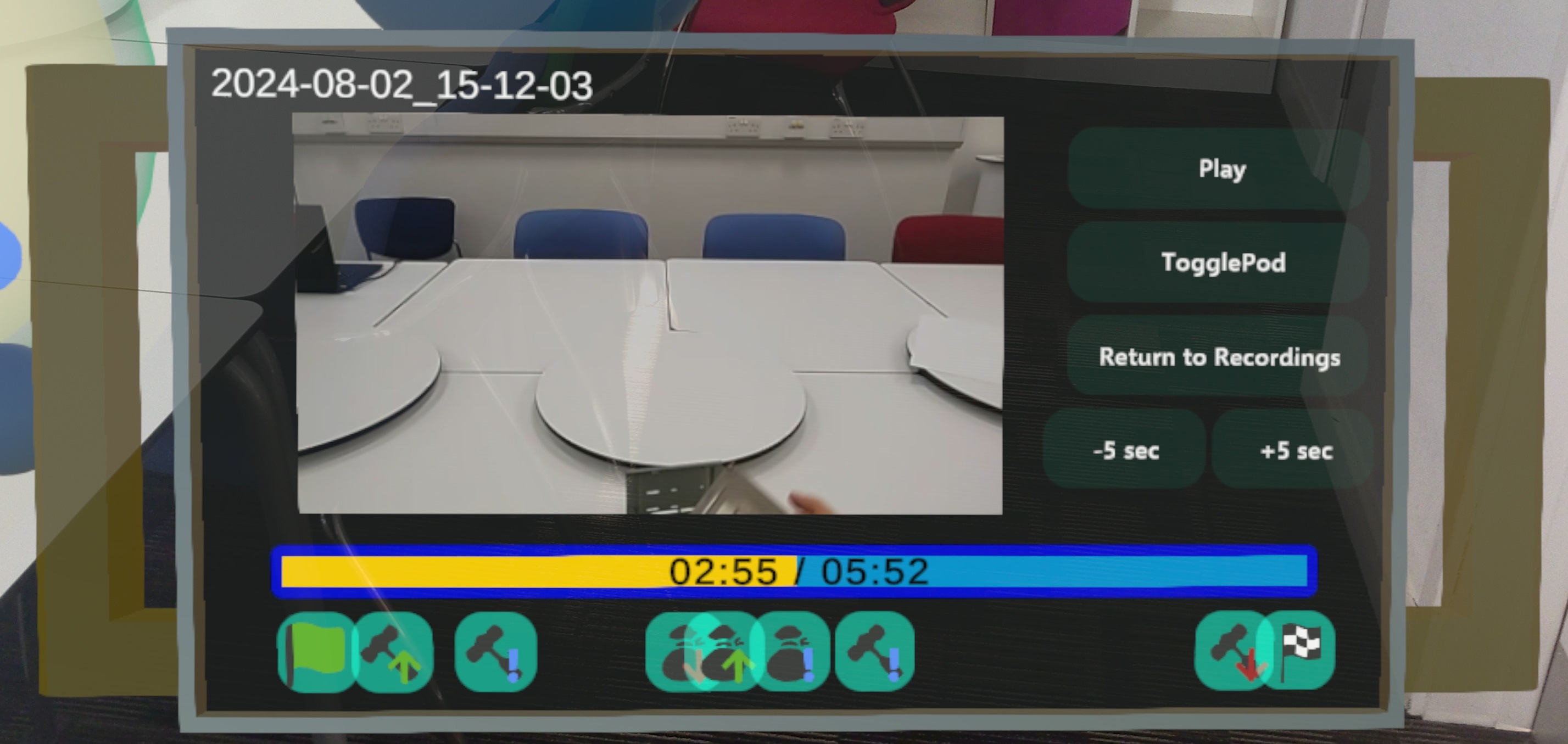}
  \caption{Keyframe navigation interface in the virtual device. Users can navigate recordings by utilising keyframes based on annotated events. These keyframes allow users to efficiently jump to significant moments and interactions within the recording, facilitating precise and targeted review.}
  \label{fig:keyframe_navigation}
 
\end{figure}

Across all replay options, keyframes deduced by the contextual annotations of the recording, are provided in the virtual device, allowing to quickly navigate critical moments, such as significant interactions, important decision points, or notable changes, as demonstrated in Figure \ref{fig:keyframe_navigation}. For example, in a collaborative remote maintenance scenario, keyframes might highlight critical steps in the maintenance process or moments when specific parts were identified as needing replacement. This allows users to quickly revisit these pivotal points during the replay, ensuring that key issues and procedural instructions are clearly understood and effectively addressed. Whether the event is being reviewed in real-scale, or in a miniature scale, these keyframes provide a seamless way to jump directly to essential moments, thereby enhancing the efficiency of the review process. 

\section{Evaluation}

\begin{figure}[t!]
  \centering  
  \includegraphics[width=0.85\linewidth]{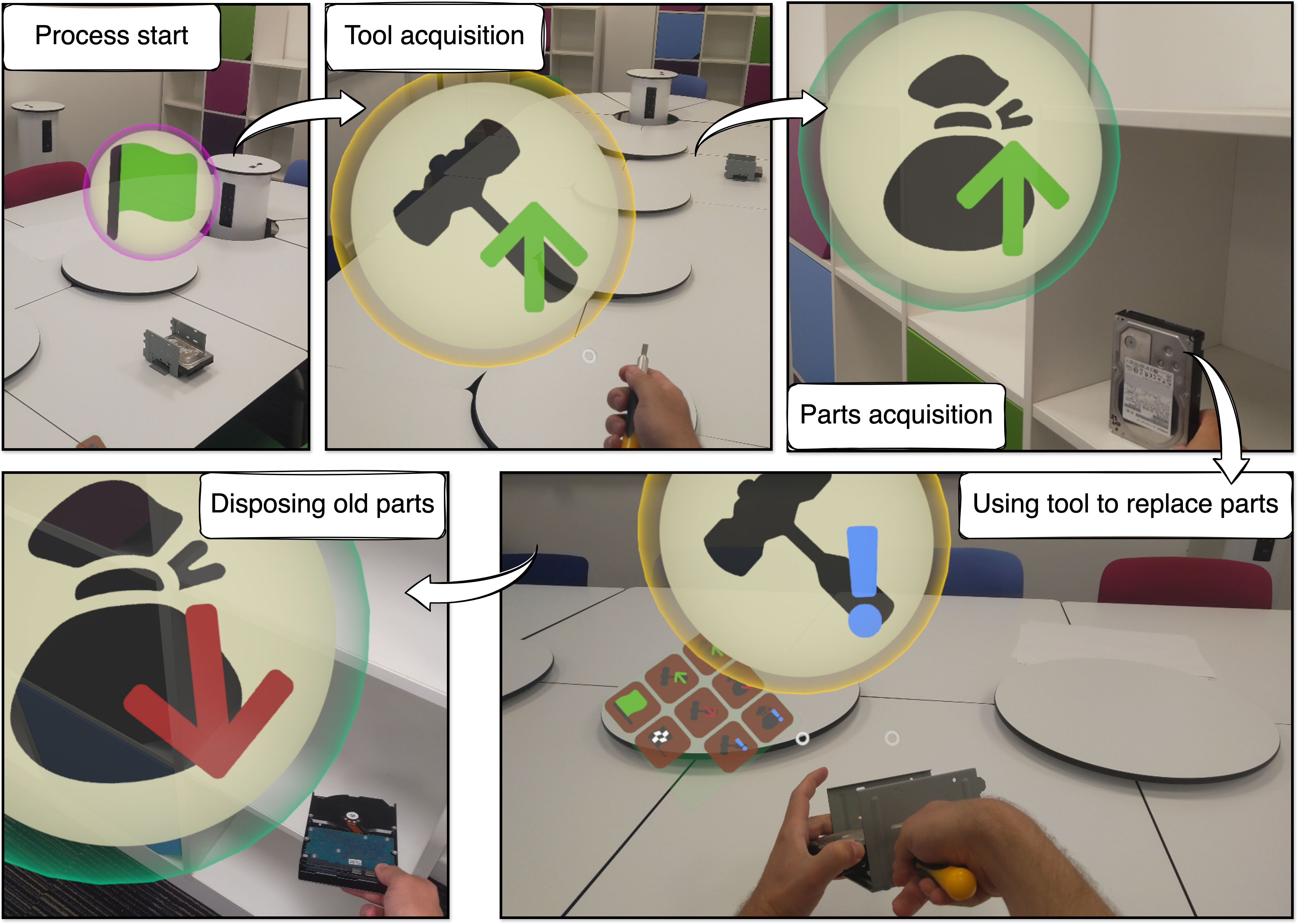}
  \caption{The flow of the maintenance process of replacing a hard drive, during the XR recording.}
  \label{fig:process_flow}
 
\end{figure}

An evaluation study was conducted in a controlled laboratory environment of a collaborative remote maintenance procedure. Specifically, the scenario involved the task of updating a hard drive within a system rack. Figure \ref{fig:process_flow} depicts the flow of the maintenance process, during the XR recording. The primary objective of the study was to assess participants' ability to accurately identify, understand, and recall the key components and steps of the maintenance process. Participants were provided with varying modes of instruction, ranging from traditional textual descriptions to dynamic XR recordings created using the proposed system. The study focused on evaluating how effectively each method conveyed essential information, particularly regarding the identification of key events within the maintenance procedure, such as the position, duration, timing, and location of critical tasks and components. The research questions aimed to explore how different presentation formats impacted participants' comprehension, retention, and ability to navigate the maintenance environment, with a specific interest in the advantages offered by the XR system in enhancing these aspects. 

\subsection{Procedure}
The experiment involved 20 participants, with a gender distribution of 15 male, 4 female, and 1 non-binary, ranging in age from 18 to 44 years. Four groups of 5 participants were defined, with each group exposed to distinct forms of instructional media. Two groups were assessed using the XR device, one based on a reconstruction in the actual environment it was recorded, and one based on the miniature scale version, providing an immersive experience of the maintenance process based on the proposed system. The other two groups relied on traditional methods: one group received text-based instructions, while the other viewed a video-based walkthrough of the procedure. Prior to the evaluation, participants in each group were given detailed explanations and source materials relevant to their assigned instructional method, to ensure they were adequately informed about the maintenance procedure. Notably, 80\% of the participants who used the XR device had prior experience with XR technology, which could potentially influence their interaction and performance within the system.

\subsection{Measures}
The evaluation was designed to measure two primary aspects of participants' performance: temporal and spatial accuracy, as defined below:

\textbf{Temporal accuracy} refers to the participants' ability to correctly identify and sequence the key events in the maintenance procedure within the temporal framework. Participants were asked to recount the specific time on which specific events occurred. This metric was particularly crucial in assessing how well the different instructional methods conveyed the timing and order of actions required for the hard drive update. 

\textbf{Spatial accuracy} evaluates the participants' ability to correctly locate and identify the critical components and areas within the maintenance environment. A detailed floor plan of the maintenance area was provided to each participant, highlighting specific zones where key actions occurred. Participants were required to map out the steps of the procedure onto the floor plan, identifying the exact locations where each task should be performed. This metric helped determine how effectively each instructional method facilitated spatial understanding of the environment.

Additionally, the usability of the different instructional methods was assessed using the standard System Usability Scale (SUS). The provided questionnaire consisted of 10 statements that participants rated on a 5-point Likert scale, ranging from "Strongly Disagree" to "Strongly Agree." The SUS was employed to capture participants' subjective perceptions of ease of use, efficiency, learnability, and overall satisfaction with the system being evaluated.

\subsection{Results}

\begin{table}[t!]
    \caption{Comparison of key metrics across different interaction modalities (Text, Video, XR Miniature Variant, and XR Real Scale). The table presents the evaluation process length, average time offset from expected, area identification accuracy, and System Usability Scale (SUS) scores. The XR Real Scale modality demonstrated the highest usability, with the shortest evaluation time, the smallest time offset, perfect area identification accuracy, and the highest SUS score. In contrast, the Text modality showed the lowest SUS score and was the fastest but lacked data for time offset and area identification accuracy.}
	\label{tab:results}
	  \begin{tabularx}{\linewidth}{ccccc}
		\toprule
		Metric & Text & Video & XR Miniature Var. & XR Real Scale \\
		\toprule
        Evaluation Process Length & 6 min 19 s & 12 min 46 s & 8 min 27 s & 8 min 5 s \\
        Average Time Offset From Expected & N/A & 84.68 s & 3.60 s & 2.24 s \\
        Area Identification Accuracy & N/A & 88 \% & 92 \% & 96 \% \\
        SUS Score & 55 & 77 & 81 & 97 \\ 
		\bottomrule
	  \end{tabularx}   
\end{table}

\begin{figure}[b!]
  \centering  
    \begin{minipage}[b]{0.6\textwidth}
    \includegraphics[width=0.95\textwidth]{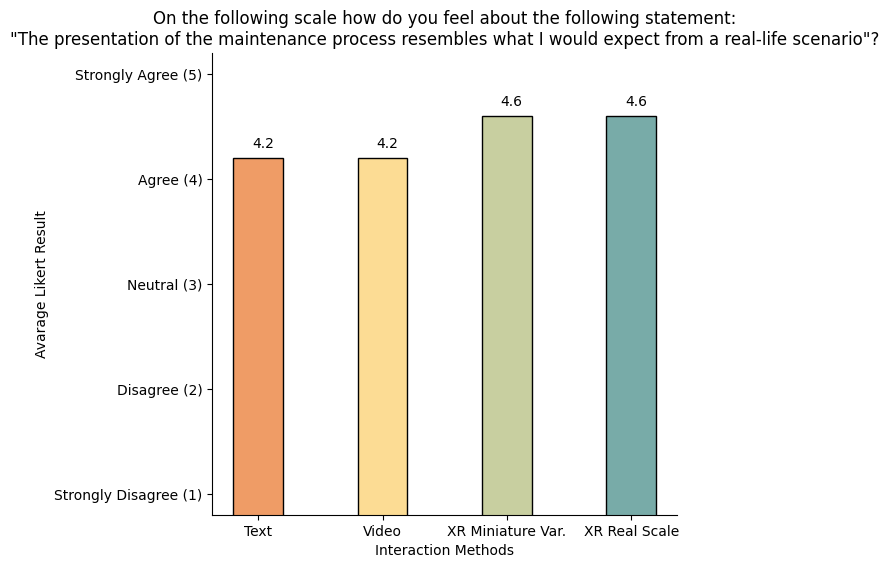}
    \vspace{1cm}
    \caption{Average participants rating related to their evaluation of which degree the presentation of the maintenance process correlates with a real-life scenario on a 5-point Likert scale. Text- and Video-based interaction methods were evaluated at 4.2, while the XR-based method received the score of 4.6.}
    \label{fig:is_depiction_correlated _with_reality_average_barchart}
  \end{minipage}
  \hfill
  \begin{minipage}[b]{0.35\textwidth}
    \includegraphics[width=0.95\textwidth]{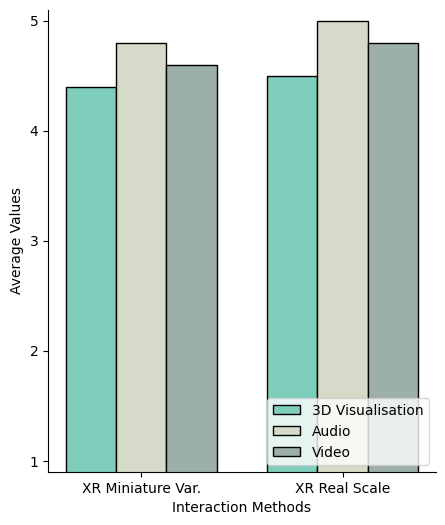}    
    \caption{Participant ratings of the perceived usefulness of different modalities in XR-based interaction methods, with all modalities scoring above 4.4 on a 5-point Likert scale. Audio was rated as the most useful, followed by video and 3D visualization.}
    \label{fig:interaction_methods_score}
  \end{minipage}
\end{figure}

The evaluation process, excluding the time allocated for viewing the source material, took on average approximately 8 minutes and 54 seconds. The average usage time for the XR device across both scenarios was slightly lower, at 8 minutes and 16 seconds. Importantly, the total interaction time with the XR device, including the familiarisation phase, did not exceed 25 minutes for any participant, and no reports of fatigue or discomfort were received during or after the evaluation process.

Participants were also asked to evaluate each method's ability to accurately represent a maintenance process in line with real-life scenarios. On average, participants rated the methods 4.4 out of 5 on a Likert scale, indicating strong agreement that the presented methods effectively mirror real-world maintenance practices. Those using the XR device predominantly expressed "Strongly Agree" responses, with an average score of 4.6, while the Video-based and Text-based groups both rated the methods slightly lower, with average scores of 4.2. The results pertaining to the depiction of the maintenance process are shown in Figure \ref{fig:is_depiction_correlated _with_reality_average_barchart}. Figure \ref{fig:interaction_methods_score} illustrates participants' perceptions of the usefulness of different modalities in the XR-based interaction methods. The consensus indicated that all modalities were perceived as highly important, with audio perceived as slightly more useful, followed closely by video, and 3D visualisation.

The results of the evaluation, including the duration of the process are outlined in Table \ref{tab:results}. The assessment of temporal accuracy in identifying events, highlighted the effectiveness of the provided spatio-temporal annotations in supporting the review process of the maintenance scenario. Both XR-based methods demonstrated relatively high accuracy in assessing the timing of events. When using the Miniature Scale version, participants selected an average offset of 2.24 seconds from the expected times, while when using the Real Scale representation, the offset was slightly higher, at 3.6 seconds. In contrast, the Video-based method, which did not feature specific event indicators and relied only on participants' personal estimates and recall, resulted in a significantly higher average offset of 84 seconds. The largest contributor to this discrepancy was the misestimation of the maintenance process duration, which presented an average difference of 130 seconds from the actual duration. Excluding this factor reduced the average time difference for the video-based method to 15.4 seconds, which still represented a substantial variance.

\begin{figure}[t!]
  \centering  
  \includegraphics[width=\linewidth]{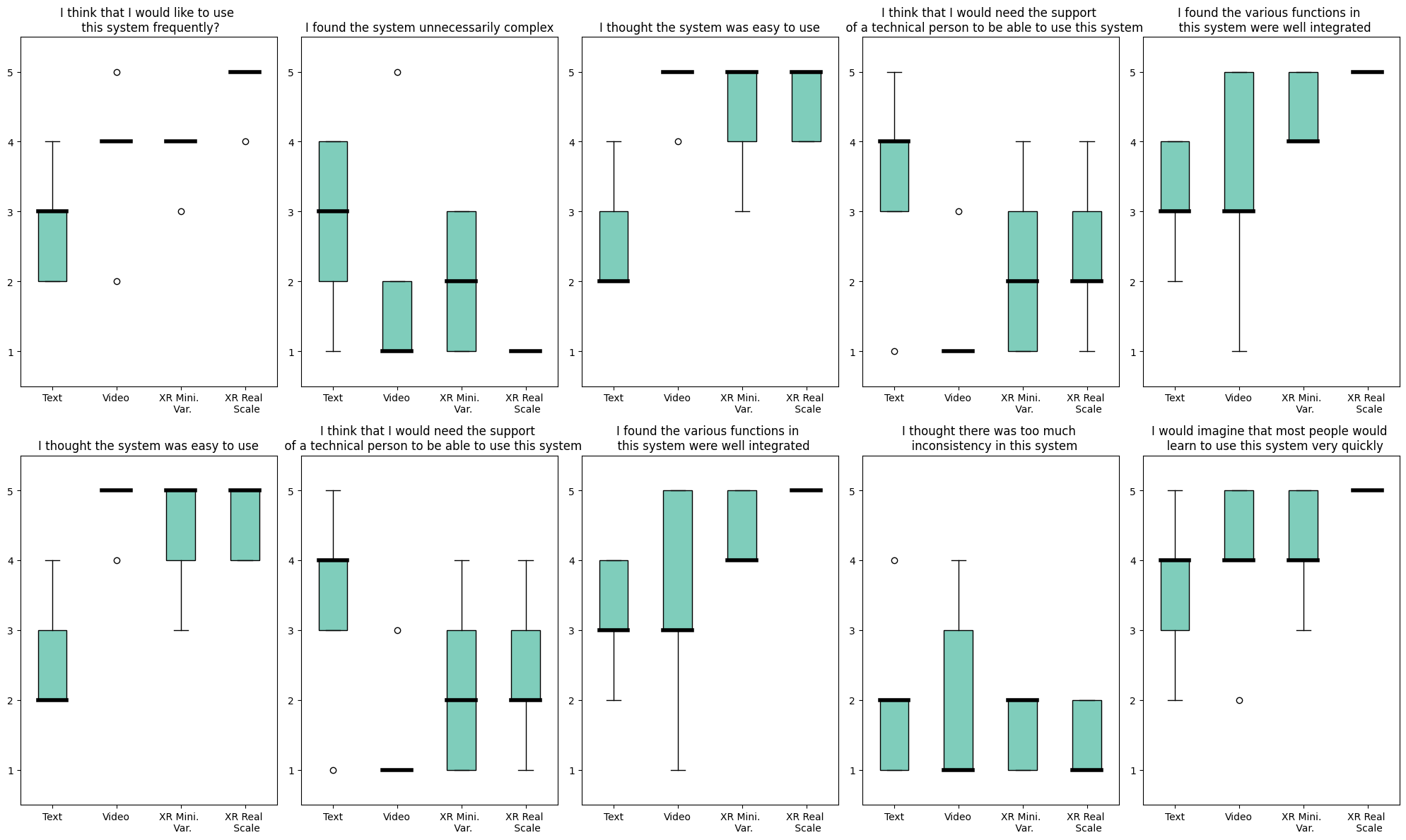}
  \caption{Box plots summarising participants' responses to the usability study across different interaction modalities (Text, Video, XR Miniature Variant, and XR Real Scale). The plots illustrate ratings on a 5-point Likert scale for various aspects of system usability, including frequency of use, perceived complexity, ease of use, need for technical support, integration of functions, and overall consistency. The XR Real Scale modality received the highest ratings, particularly in terms of ease of use and integration, while the Text-based method was generally rated lower across multiple dimensions.}
  \label{fig:usability_study}

\end{figure}

Regarding the spatial accuracy in identifying specific areas where maintenance tasks were performed, the XR Real Scale modality enabled participants to achieve 96\% accuracy, followed by the XR Miniature Variant with 92\%. In contrast, the video-based method achieved slightly less accuracy of 88\%. The findings demonstrate that XR modalities, particularly the Real Scale variant, significantly improve spatial accuracy in identifying maintenance areas, outperforming traditional video-based methods. The slight difference between the Real Scale and Miniature Variant suggests that while both XR approaches enhance spatial precision, the immersive nature of full-scale XR may offer a more intuitive understanding of the environment.

One limitation of the 3D environment reconstruction which might contribute to the slightly lower accuracy of the miniature scale version, is the lack of visual detail. While the current implementation leverages depth sensors to create point clouds and meshes for spatial awareness in the XR device, the absence of color data limits the fidelity of the reconstructed environment. This limitation arises from the fact that depth sensors only measure distance, neglecting the visual information that contributes to a realistic representation. Neural Radiance Fields (NeRF) \cite{DBLP_Nerf2020, NEURIPS2023_aebf6284} offer a promising solution to this, by capturing not only depth but also color, reflections, and transparency. This approach enables the creation of highly detailed and photorealistic virtual environments. However, the computational demands and processing time currently required for NeRF-based reconstructions in real-time, hinder their practical application in XR asynchronous communications.

The usability study assessed participants' perceptions of the system's ease of use, integration of functions, and overall user experience. Participants were asked to evaluate the system on various aspects, including its complexity, the need for technical support, and their confidence in using it. The study also gauged the perceived learning curve and the consistency of the system's functions. Overall, the questions aimed to provide a comprehensive understanding of how intuitive and user-friendly the system was for the participants. 

The results of the usability study are depicted in Figure \ref{fig:usability_study}. The data suggests that participants generally found the XR-based methods to be more favourable compared to the Text and Video-based methods. Specifically, participants indicated a stronger preference for using the XR system frequently and found it easier to use, with fewer reports of unnecessary complexity. The XR Real Scale modality consistently received high ratings across various questions, particularly in terms of ease of use, integration of functions, and user confidence. In contrast, the Text-based method was rated lower, particularly in terms of ease of use and the likelihood of requiring technical support. These findings highlight the potential of XR-based methods to provide a more intuitive and efficient user experience compared to traditional methods.

\section{Conclusion}

This paper introduces a novel system designed to significantly enhance asynchronous communication in XR environments, with a particular focus on collaborative remote maintenance applications. The system features MemoryPods, which allow users to record, annotate, and replay interactions in an immersive manner. This capability improves both temporal and spatial precision when reviewing complex tasks. Additionally, AI-driven summarisation reduces cognitive load by generating concise overviews of key events, ensuring that critical information remains easily accessible. Together, these features address important challenges in asynchronous XR communication, offering users efficient tools for revisiting and analysing past interactions.

An empirical study validates the system's effectiveness and usability within a collaborative remote maintenance scenario. Participants demonstrated enhanced comprehension and more accurate task recall when using the MemoryPod system compared to traditional text and video-based methods. These findings suggest that the system has the potential to transform workflows, not only in remote maintenance but also in fields such as education, training, and healthcare, by offering a more immersive and interactive experience. Although certain limitations exist, such as the visual fidelity of 3D reconstructions, ongoing advancements in XR and AI technologies are likely to further improve the system’s functionality.

\bibliographystyle{ieeetr}
\bibliography{ref}

\end{document}